\documentclass[prc,twocolumn,showpacs,showkeys,superscriptaddress]{revtex4}

\usepackage{graphicx}
\usepackage{amsfonts}
\usepackage{amsmath}
\usepackage{amssymb}

\newcommand{\dbar}{\raisebox{0.44em}
                  {\rule{1ex}{0.04em}}\hspace{-1.2ex}\lambda_N}

\begin{document}

\title{Laser-induced nonresonant nuclear excitation in muonic atoms}

\author{A.\ Shahbaz}
\altaffiliation{Permanent address: Department of Physics, GC University, 54000 Lahore, Pakistan}
\affiliation{Max-Planck-Institut f\"{u}r Kernphysik, Saupfercheckweg 1, 69117 Heidelberg, Germany}
\author{C.\ M\"{u}ller}
\affiliation{Max-Planck-Institut f\"{u}r Kernphysik, Saupfercheckweg 1, 69117 Heidelberg, Germany}
\author{T.\ J.\ B\"{u}rvenich}
\affiliation{Frankfurt Institute for Advanced Studies, Johann Wolfgang Goethe University, Ruth-Moufang-Str. 1, 60438 Frankfurt am Main, Germany}
\author{C.\ H.\ Keitel}
\affiliation{Max-Planck-Institut f\"{u}r Kernphysik, Saupfercheckweg 1, 69117 Heidelberg, Germany}

\date{\today}

\begin{abstract}
Coherent nuclear excitation in strongly laser-driven muonic atoms is calculated. 
The nuclear transition is caused by the time-dependent Coulomb field of the oscillating charge density of the bound muon. A closed-form analytical expression for electric multipole transitions is derived and applied to various isotopes; the excitation probabilities are in general very small. We compare the process with other nuclear excitation mechanisms through coupling with atomic shells and discuss the prospects to observe it in experiment.
\end{abstract}



\maketitle

%
\section{Introduction}
Excitation of atomic nuclei has been one of the major subjects to be investigated by physicists  for almost a century. Various mechanisms are capable to change the nuclear quantum state \cite{EiGr1970}. In particular, transitions between atomic shells can couple to nuclear degrees of freedom. For example, when the energetic difference between two atomic states matches a low-lying nuclear transition energy ($\hslash\omega_N\lesssim 100$\,keV), the energy released during the atomic deexcitation can be transfered resonantly to the nucleus leading to its excitation (nuclear excitation by electron transition, NEET) \cite{NEET,NEET2}. Similar mechanisms proceed via electron capture or scattering \cite{NE,NEEC,muonicNEEC}. Despite their rather small probabilities, these kind of processes are of both fundamental and practical interest since potential applications comprise the efficient triggering of isomeric nuclear states \cite{AdJoCh2007} and especially the development of a nuclear $\gamma$-ray laser \cite{GRASER}.

At the borderline between atomic and nuclear physics, muonic systems play a prominent role \cite{MuRev}. Because of the small Bohr radius of the bound muon, there is an appreciable influence of the nuclear structure on the atomic states and vice versa. Muonic atoms therefore represent powerful tools for nuclear spectroscopy for more than 50 years. In fact, while NEET has been measured for the first time in $^{189}$Os in the mid 1970s \cite{NEETexp}, with conclusive evidence even only recently in $^{197}$Au \cite{NEETexp2}, in muonic atoms the equivalent process was already observed in 1960 \cite{muonicNEETexp,muonicNEET1,muonicNEET2,WiDaSe1954}. Today, facilities like TRIUMF (Vancouver, Canada) or PSI (Villingen, Switzerland) are specialized in the efficient production of muons and muonic atoms \cite{TRIUMF}. Apart from nuclear excitation, bound muons are also able to catalyze nuclear fission \cite{Ga1980} and fusion \cite{BrKaCoLe1989} reactions. Recent developments aim at the generation of beams of radioactive muonic atoms for pursuing spectroscopic studies on unstable nuclear species \cite{MuRadioactive}.

Strong laser fields can serve as an additional bridge between atomic and nuclear physics \cite{Matinyan,reviews}. It has been proposed, for example, that in laser-assisted NEET \cite{KaKe1993} a possible mismatch between the atomic and nuclear transition energies can be compensated by the simultaneous emission or absorption of laser photons. Nuclear excitation via rescattering of field-ionized electrons has also been studied theoretically \cite{Mocken,recollision1,recollision2}. With the advent of intense x-ray free-electron laser (XFEL) sources even direct laser-nucleus coupling via coherent photoexcitation may become possible \cite{1BuEvKe2006}. Experiments on intense laser-plasma interactions have already observed incoherent nuclear reactions through laser-generated electrons and bremsstrahlung \cite{ScMaBe2006}. At present the most powerful laser sources reach intensities of $\sim 10^{22}$ W/cm$^{2}$ in the near-infrared frequency domain \cite{Yanovsky}, and an increase to $\sim 10^{26}$ W/cm$^{2}$ is envisaged \cite{ELI}. Similar intensities might be attainable with XFEL radiation \cite{XFEL}. Such superintense laser beams can influence the dynamics of bound muons because they are comparable in strength with the Coulomb fields in light muonic atoms. These correspond to $\approx 4\times 10^{25}$\,W/cm$^{2}$ in the ground state of muonic hydrogen and raise like $Z^6$ with atomic number. So far, the interaction of intense laser fields with muonic atoms and molecules (H, D, and D$_2^+$) has been considered with respect to nuclear fusion \cite{ChBaCo2004}, dynamical nuclear probing \cite{ShMuStBuKe2007}, and the Unruh effect \cite{KaSc2005}. It is important to note that light muonic atoms are practically stable on the ultrashort time scale of strong laser pulses ($\tau \sim$ fs$-$ns) since the free muon life time amounts to $2.2\,\mu$s. In heavy muonic atoms, muon absorption by the nucleus reduces the lifetime of deeply bound states significantly.

In this paper, we calculate coherent nuclear excitation in hydrogenlike muonic atoms which are exposed to superintense laser fields. We restrict the consideration to nuclear charges $Z\lesssim 10$ since otherwise the required laser intensites become unrealistically large. Driven by the field, the bound muonic charge cloud oscillates  periodically across the nucleus leading to electromagnetic nuclear excitation (see Fig.~1). In contrast to NEET, this effect does not rely on a resonance condition. It has been studied before in electronic atoms \cite{SoBi1988,BeGoWe1991,Hartmann,innershell}, with a focus on the transition to the very low-lying isomeric level in $^{235}$U at 76\,eV, but the predicted excitation probabilities are small and could not yet be verified in experiment \cite{{BoDy1992}}. From the experimental data an upper bound for the excitation probability of $\sim 10^{-5}$ was extracted. We point out that contrary to laser-generated plasma experiments \cite{ScMaBe2006}, the nucleus is excited solely by its own electron or muon. The process might be called nuclear excitation by coherent electron (muon) motion, NECEM (NEC$\mu$M). We show that muonic atoms are in principle favorable candidates to observe the effect as the muon produces a much higher charge density within the nuclear volume. The excitation probabilities are always very small, though, because the driving laser frequency is far off resonance with the nuclear level spacing. We present calculations for nuclear electric multipole transitions as a function of the applied laser frequency and intensity, and discuss the possibility to detect the effect by suitable experimental arrangements. Our results indicate that observation of NEC$\mu$M represents a challenging task, but it might come into experimental reach by powerful XFEL facilities in the near future.

\begin{figure}
\includegraphics*[width=0.9\linewidth,angle=0]{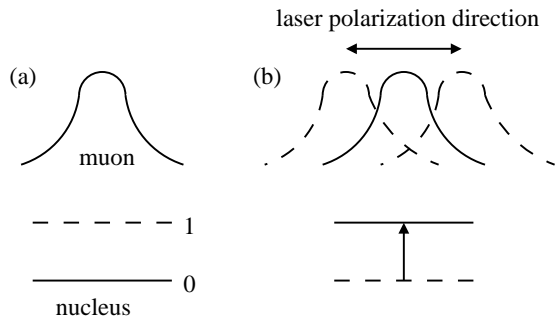}
\caption{Schematic diagram of nuclear excitation by coherent muon motion, NEC$\mu$M. (a) Initially, in the absence of an external field, the muon and nucleus of the hydrogenlike muonic atom are in their respective ground states. (b) When exposed to a strong laser field, the muonic charge cloud is driven into oscillation which leads to nuclear excitation $|0\rangle\to |1\rangle$ via the resulting time-dependent Coulomb interaction.}
\label{scheme}
\end{figure}

%
\section{Outline of the model}
We assume a hydrogenlike muonic atom in the ground state which is exposed to an external laser field. The combined influence of the nuclear Coulomb field and the laser field on the muon produces a time-dependent charge density $\rho({\bf r},t)=e^2|\psi({\bf r},t)|^2$, with the muon wave function $\psi$, which can lead to excitation of the nucleus. The Hamiltonian for the interaction between the muonic and nuclear charge densities is given by \cite{EiGr1970}
\begin{eqnarray}
\label{H}
H_{\rm int}(t) = \int d^3r \int d^3r_N \frac{\rho({\bf r},t)\rho_N({\bf r}_N)}{|{\bf r}-{\bf r}_N|}\,.
\end{eqnarray}
The nuclear long-wavelength limit has been applied here because the nuclear $\gamma$-ray wavelength $\dbar= c/\omega_N\approx 10^3\,{\rm fm}$ of low-lying transitions is much larger than the nuclear and atomic extensions ($r$, $r_N\lesssim 10$\,fm). Moreover, since we are interested in electric multipole transitions we neglected the current-current part which would give rise to magnetic transitions. After a multipole expansion of the Coulomb interaction in Eq.\,(\ref{H}), the probability for an electric transition between the nuclear states $|0\rangle$ and $|1\rangle$ becomes (within the first order of perturbation theory)
\begin{eqnarray}
\label{P01}
P_{0\rightarrow1}(E\ell)=\left(\dfrac{4\pi e}{\hslash}\right)^{2} \dfrac{B(E\ell)}{(2\ell +1)^{3}}\left|\int_{0}^{T} dt\ F_\ell(t) e^{i \omega_{N}t} \right|^{2}
\end{eqnarray}
with the laser pulse duration $T=2\pi N/\omega$, the nuclear transition energy $\hslash\omega_N$, the multipolarity $\ell$, and the function \cite{BeGoWe1991}
\begin{eqnarray}
\label{F}
F_\ell(t) = \int d^3r \dfrac{\rho(\textbf{r},t)} {r^{\ell+1}}Y_{\ell}^0(\Omega)\,.
\end{eqnarray}
Here, $Y_{\ell}^0(\Omega)$ is a spherical harmonic and cylindrical symmetry along the $z$ axis is employed. The reduced transition probability $B(E\ell)$ in Eq.\,(\ref{P01}) results from the integral of the nuclear transition density $\rho_N$ over nuclear coordinates in the usual way. Our goal is to obtain a closed-form analytical expression for $P_{0\to 1}(E\ell)$. To this end we follow the model developed in Ref.\,\cite{BeGoWe1991}. We apply the dipole approximation to the laser field ${\bf E}(t)=E_0\sin(\omega t){\bf e}_z$ which couples to the muon via $e{\bf E}\cdot{\bf r}$. This approximation is well justified since the muonic Bohr radius $a_0$ is much smaller than the laser wave length, and the laser field strength will be restricted to values where the muon dynamics stays nonrelativistic. Moreover, the binding Coulomb field of the nucleus is modeled by a spherical harmonic oscillator potential $V(r) = \frac{1}{2} m \omega_0^2 r^2$, with the muon mass $m$, the oscillator frequency $\omega_0$, and the muon radial coordinate $r$ \cite{BeGoWe1991,Taylor}. This procedure has proven useful for a nonperturbative, though approximate description of the laser-driven dynamics of bound states which otherwise is impossible by analytical means. When a particle is bound in a harmonic potential, the influence of an external laser field can be taken into account to all orders [see Eq.\,(\ref{psi}) below]. It is interesting to note that in the case of muonic atoms the approximation becomes the better the heavier the binding nucleus is because the potential inside an extended nucleus (considered as a homogenously charged sphere) is indeed harmonic. In very heavy muonic atoms the orbital radius is so small that the muon spends much of its time in the nuclear interior. However, as motivated before, we will consider light muonic systems where the use of the harmonic oscillator potential clearly represents an approximation only. We stress that this approach is exploited solely to treat the field-induced time evolution of the atomic state; the muonic interaction with the nucleus is correctly described by a Coulomb potential [see Eq.\,(\ref{H})]. The oscillator length $a_0=\sqrt{\hslash/m\omega_0}$ is chosen to coincide with the atomic Bohr radius. With this choice, also the actual binding energy agrees with the oscillator ground-state energy. In Fig.\,2 the harmonic oscillator ground state is compared with the Coulombic 1s state. Due to the rather similar shapes of the densities, this method allows an order-of-magnitude estimate of $P_{0\to 1}(E\ell)$. The main physical implication of the approximation is that the muon cannot be ionized since a harmonic oscillator has bound states only. Being interested in nuclear excitation by the laser-driven bound muon dynamics, we therefore restrict the laser intensity to values where ionization may safely be ignored. Nuclear excitation by rescattering of ionized electrons in laser fields has been discussed elsewhere \cite{recollision1,recollision2,Mocken}. Moreover, we consider only the nonresonant case where $\omega_0$ is significantly different from $\omega_N$. In this situation the correct atomic level structure is of minor importance. (Note that for $\omega_0\approx\omega_N$ the NEET process is possible anyway.)

\begin{figure}
\includegraphics*[width=0.9\linewidth,angle=0]{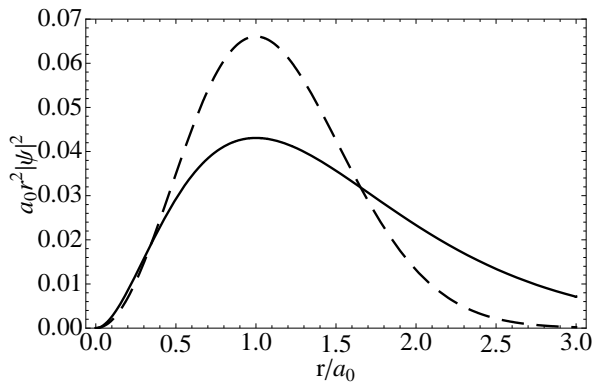}
\caption{Comparison of the ground state wave functions in a Coulomb (solid line) and a harmonic oscillator potential (dashed line). Shown is the square modulus weighted by $r^2$, with $a_0$ denoting the Bohr radius and the oscillator length, respectively. Both wave functions coincide in their characteristic properties having the same energy eigenvalues and radial peak positions.}
\label{ground}
\end{figure}

The Schr\"odinger equation for the muon motion in the combined fields can be solved analytically. Up to an irrelevant phase factor, the muon wave function reads \cite{BeGoWe1991}
\begin{eqnarray}
\label{psi}
\psi({\bf r},t) = \phi({\bf r}-{\bf u}(t))
\end{eqnarray}
where $\phi(r)=(1/\pi a_0^2)^{3/2}e^{-r^2/a_0^2}$ is the ground-state wave function in the harmonic oscillator potential and ${\bf u}(t) = u_0[\sin(\omega t)-(\omega/\omega_0)\sin(\omega_0 t)]{\bf e}_z$ is the periodic displacement caused by the laser field, with $u_0 = eE_0/m(\omega_0^2-\omega^2)$. In the limit $\omega\ll\omega_0$ of interest here, ${\bf u}(t) \approx (eE_0/m\omega_0^2)\sin(\omega t){\bf e}_z$ looks similar to the classical trajectory of a free muon in the laser field, but the excursion amplitude is reduced by a factor $(\omega/\omega_0)^2$ due to the harmonic binding force. Eq.\,(\ref{psi}) has an intuitive interpretation of the muon time evolution: the wave packet keeps its shape but is periodically shifted across the nucleus by the driving laser field. With the corresponding charge density $\rho({\bf r}-{\bf u}(t))$, the spatial integral in Eq.\,(\ref{F}) can be solved exactly. For electric dipole, quadrupole and octupole transitions we obtain
\begin{eqnarray}
\label{F123}
F_1(t) &=& \frac{\sqrt{3}}{2\pi u^2}\left[ \sqrt{\pi}{\rm Erf}(x) - 2x e^{-x^2} \right] \nonumber\\
F_2(t) &=& \frac{\sqrt{5}}{2\pi u^3}\left[ \sqrt{\pi}{\rm Erf}(x) - \left( 2x + \frac{4}{3}x^3\right) e^{-x^2} \right] \\
F_3(t) &=& \frac{\sqrt{7}}{2\pi u^4}\left[ \sqrt{\pi}{\rm Erf}(x) - \left( 2x + \frac{4}{3}x^3+\frac{8}{15}x^5\right) e^{-x^2} \right] \nonumber
\end{eqnarray}
with the Gaussian error function ${\rm Erf}(x)$ and $x=u(t)/a_0$. In order to prevent field-induced ionization, the laser electric field strength should be far below the barrier-suppression value $E_{\rm OBI}=(\alpha Z)^3m^2c^3/16e\hslash$, where the binding Coulomb potential is suppressed by the laser field all the way to the bound energy level \cite{reviews}. This implies $u_0\ll a_0$ (in fact, $u_0=a_0/16$ at $E_0=E_{\rm OBI}$) and, moreover, the muon velocity $v\sim\omega u_0\ll c$ remains nonrelativistic. We may therefore perform in Eq.\,(\ref{F123}) a Taylor series expansion in the small parameter $x$ yielding $F_\ell(t) \sim u(t)^\ell$. The time integral in Eq.\,(\ref{P01}) then gives 
\begin{eqnarray}
\label{P01final}
P_{0\to 1}(E\ell) \approx C_\ell\, \alpha^2 \frac{B(E\ell)}{e^2a_0^{2\ell}}\, \frac{\dbar^2}{a_0^2} \left(\frac{u_0 \omega}{a_0 \omega_N}\right)^{2\ell} \,,
\end{eqnarray}
with the QED fine-structure constant $\alpha=\frac{e^2}{\hslash c}$, the numerical constants $C_1=\frac{128}{81}=1.58$, $C_2=\frac{2048}{5625}=0.364$, and $C_3=\frac{8192}{60025}=0.136$ for electric dipole, quadrupole and octupole transitions, respectively, and assuming $\omega_0 > \omega_N$ so that the second term in ${\bf u}(t)$ proportional to $\sin(\omega_0t)$ may be ignored. We note that a fast oscillating factor $\sin^2\left(\frac{1}{2}\omega_N T \right) = \sin^2\left(\pi N\omega_N /\omega\right)$ occuring in Eq.\,(\ref{P01}) when evaluating the time integral has been averaged over frequency in Eq.\,(\ref{P01final}) to produce a factor $\frac{1}{2}$. The reason is that an intense short laser pulse comprises a large frequency bandwidth so that the sin$^2$ term will adopt a different value for each spectral component. A sin$^2$ time dependence is typical for the population dynamics of a two-level system in an external periodic field \cite{1BuEvKe2006,SoBi1988}.

The nuclear excitation probability in Eq.\,(\ref{P01final}) essentially scales like $P_{0\to 1}(E\ell)\propto a_0^{-2(\ell +1)}$ with the Bohr radius, which clearly demonstrates the expected result that compact atomic states are advantageous. For example, the probability for a nuclear $E1$ transition in a hydrogenlike muonic atom is larger by 9 orders of magnitude than in the corresponding electronic system, when the laser field strength is accordingly scaled so that the ratio $u_0/a_0$ is identical. Note that $B(E\ell)\propto e^2 R_N^{2\ell}$ with the nuclear radius $R_N$ so that a factor $(R_N/a_0)^{2\ell}$ is contained in Eq.\,(\ref{P01final}). Apart from this scaling, the atomic size enters through the factor $(u_0/a_0)^{2\ell}$ which depends on the applied laser intensity. The appearance of the ratio $u_0/a_0$ is intuitive since the larger its value the closer the muon comes to the nucleus, this way increasing the mutual Coulomb interaction. By chosing appropriately large laser fields with $E_0\lesssim E_{\rm OBI}$, the ratio $u_0/a_0$ can be optimized to values of several percent. Via the displacement $u_0$, the excitation probability depends like $P_{0\to 1}(E\ell)\propto E_0^{2\ell}$ on the laser field strength. This behaviour is reminiscent of multiphoton processes in atoms or molecules which scale as $E_0^{2n}$ when $n$ laser photons are involved and perturbation theory applies \cite{reviews}. The photon order $n$ formally corresponds to the multipolarity $\ell$ of the transition here. Within this analogy, the excitation mechanism might be interpreted as 'multiphonon' absorption from the periodically oscillating muon charge density. The main factor, however, determining the absolute value of the probability is the frequency ratio $(\omega/\omega_N)^{2\ell}$. In optical or infrared laser fields, the large frequency mismatch suppresses the nonresonant process by many orders of magnitude since the lowest transition energies in light nuclei are $\hslash\omega_N\sim 100$\,keV. 

\begin{table}[h]
\centering
\caption{Parameters of the atomic nuclei under consideration. Given are the electric multipolarity, nuclear transition energy, and reduced transition probability (in Weisskopf units). The last two columns contain the atomic binding energy and Bohr radius of hydrogen-like muonic $^{19}$F, $^{16}$N, and electronic $^{235}$U (treated nonrelativistically) in the 1s ground-state. The data are taken from \cite{NEET2,data}.}
\begin{tabular}{c|ccc||cc}
\hline
nucleus & type of & $\hslash \omega_{N}$ & $B(E\ell)$ & $\frac{1}{2}\hslash 
\omega_{0}$  & $a_{0}$\\
& transition & [keV] & [w.u.] & [keV] & [fm]\\
\hline
$^{19}$F  & $E1$ & $110$   & $0.0012$ & $228$ & $28.4$\\
$^{16}$N  & $E2$ & $120$   & $1.7$    & $138$ & $36.5$\\
$^{235}$U & $E3$ & $0.076$ & $0.0007$ & $115$ & $575$\\
\hline
\end{tabular}
\label{table2}
\end{table}

%
\section{Results and discussion}
Based on Eq.\,(\ref{P01final}) we have calculated the NEC$\mu$M probability in hydrogenlike muonic $^{19}$F and $^{16}$N. These nuclei possess the lowest-energy electric transitions among isotopes with $Z\lesssim 10$. We compare our results with the corresponding ones for hydrogenlike electronic $^{235}$U where the isomeric level at 76 eV is reached by an $E3$ transition. This nucleus has been studied most intensively in the literature \cite{SoBi1988,BeGoWe1991,Hartmann}. In the spirit of the present model, the $^{235}$U system is treated nonrelativistically, although in highly charged ions relativistic effects exist. The main parameters of the systems under consideration are summarized in Table I. The atomic binding energy is of the order $\sim 100$\,keV in all cases. The atomic extension of the $^{235}$U ion is larger by an order of magnitude than the muonic Bohr radii because of the smaller electron mass.

Figure~3 illustrates the laser intensity dependence of the nuclear excitation probability in Eq.\,\eqref{P01final}. Strongly laser-driven hydrogenlike muonic $^{19}$F, muonic $^{16}$N and electronic $^{235}$U are considered here, representing examples for electric dipole, quadrupole and octupole transitions, respectively. Since, for NECEM/NEC$\mu$M, the particle should stay bound in order to influence the nucleus while oscillating across it, we restrict the curves to intensities below the OBI limit. Within this range of interaction, an increase of the driving laser intensity leads to enhanced nuclear transition probabilities in accordance with the power-law dependence displayed in Eq.\,\eqref{P01final}. The increase is the steeper, the higher the multipolarity of the transition is. The largest excitation probabilities are obtained from the dipole transition in muonic $^{19}$F. The comparison of the three different systems thus highlights the relative advantage of using light muonic atoms over electronic heavy ions. The absolute values of the nuclear excitation probability are always very small, though.
The main reason for the suppression is the largely nonresonant character of the process, leading to a very small ratio $\omega/\omega_N$ in Eq.\,\eqref{P01final} which is of the order 10$^{-5}$ for the two muonic atoms considered. In other words, the time scale of the laser-driven muon motion is too slow to effectively couple to the nuclear transition. Higher laser frequencies are therefore desirable in order to reduce this detrimental mismatch.

\begin{figure}
\includegraphics*[width=0.9\linewidth,angle=0]{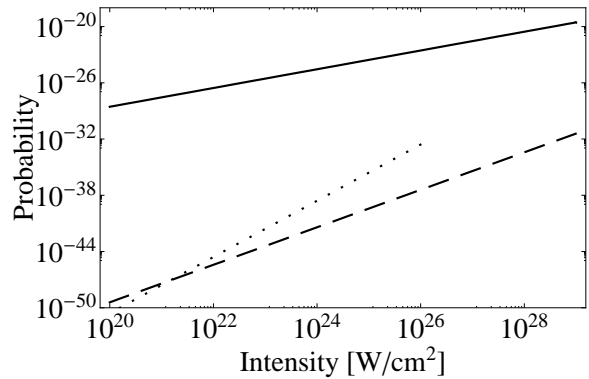}
\caption{Nuclear transition probabilities for laser-driven muonic $^{19}$F (solid curve), muonic $^{16}$N (dashed curve), and electronic $^{235}$U (dotted curve), as function of the laser intensity. The curves refer to a single atom, initially in the 1s ground-state, and stop at the respective OBI intensities. The infrared laser photon energy is $\hslash\omega = 1$\,eV. }
\label{fig1a}
\end{figure}

Figure~4 shows the dependence of the NEC$\mu$M probability \eqref{P01final} as a function of laser frequency from the infrared to the envisaged XFEL domain \cite{XFEL}. Only the muonic systems $^{19}$F and $^{16}$N are considered where $\omega\ll\omega_N$ throughout the plot range. In electronic $^{235}$U we would pass through a resonance; to this case our approach does not apply. The laser intensity amounts to $10^{26}$ W/cm$^{2}$. With increasing laser frequency the excitation probability is substantially enhanced, in particular for the non-dipole transition in $^{16}$N. The maximum probability of about 10$^{-14}$, however, is still obtained from the $E1$ transition in $^{19}$F when 10\,keV XFEL radiation is applied.
An alternative way of obtaining high laser frequencies is to employ (instead of fixed target nuclei) an ion beam which counterpropagates the laser pulse at relativistic speed.
In the nuclear rest frame the laser frequency appears Doppler-blueshifted. In this geometry even a resonant laser-nucleus coupling \cite{1BuEvKe2006} could be achieved when a bare $^{235}$U beam collides at a Lorentz factor $\gamma\approx 30$ with a near-infrared laser beam ($\hslash\omega\approx 1.2$\,eV). The Doppler-shifted laser frequency $\omega'\approx 2\gamma\omega$ can be tuned into resonance with the nuclear transition frequency. In fact, such an experiment would be taylormade for the future GSI facility where a beam of hydrogenlike or fully stripped U ions of the required energy will be available, along with the intense PHELIX laser \cite{GSI}.

\begin{figure}
\includegraphics*[width=0.9\linewidth,angle=0]{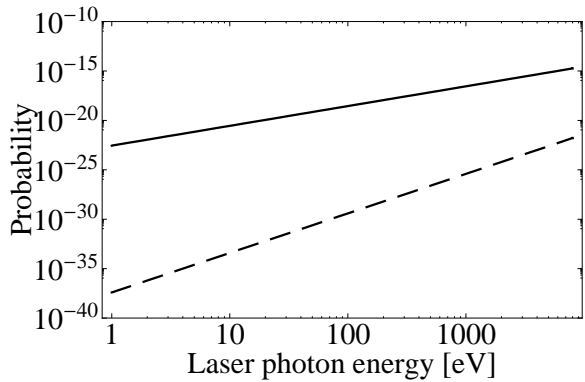}
\caption{Nuclear transition probabilities for laser-driven muonic $^{19}$F (solid curve) and $^{16}$N (dashed curve), as function of the laser frequency. The curves refer to a single atom, initially in the 1s ground-state. The laser intensity is $10^{26}$\,W/cm$^{2}$.}
\label{fig1a}
\end{figure}

Despite the very small nuclear transition probabilities shown in Figs.~3 and 4, the muon still leads to a substantial enhancement of the laser-nucleus interaction. This is clearly demonstrated by a comparison with the corresponding probability for direct nuclear excitation by the laser field. When the intensity is extremely high, nuclei can be excited directly by an off-resonant laser field  \cite{ThomasPRC}. In the present situation, however, this direct excitation channel is still of negligible importance.
Applying $n$-th order perturbation theory, the latter can be estimated as $P\sim \Gamma^{2n}$, with $\Gamma\approx eE_0R_N/\hslash\omega_N$ and $n\approx \omega_N/\omega$ \cite{Matinyan}. For the light muonic isotopes at the laser parameters $I=10^{26}$ W/cm$^{2}$ and $\hslash\omega=10$\,keV (where the maximum NEC$\mu$M probabilities are reached), we obtain $\Gamma\sim 10^{-2}$ and $n\approx 10$ so that $P\sim 10^{-40}$.

It is instructive to comment on various differences with the results on NECEM presented in \cite{BeGoWe1991}. We consider hydrogenlike atoms in the 1$s$ state which are moderately affected by an external field such that $u_0\ll a_0$. In this asymptotic limit we find simple power-law dependences of the excitation probability on the laser intensity and frequency [see Eq.\,(\ref{P01final})]. Contrary to that, in \cite{BeGoWe1991} the collective oscillation of all electrons in higher atomic shells (starting from the 2s orbital) has been considered. The driving laser field was assumed so strong that $u_0\sim a_0$. In this regime, neglecting field ionization, the excitation probability as a function of intensity first exhibits a power-law increase (with a different exponent, though) and eventually goes through a maximum. The absolute values of the excitation probability in $^{235}$U are larger than in our case: for the 4s and 5s electrons, they typically amount to $P\sim 10^{-15}$ at about $10^{23}$\,W/cm$^{2}$ laser intensity and 5\,eV photon energy, assuming $B(E3)$ to be one Weisskopf unit \cite{BeGoWe1991}.

Regarding the experimental measurability of the NEC$\mu$M process, we note that the  probabilities shown in Figs.~3 and 4 refer to a single atom. When more than one atom interacts with the laser field, the total yield could be increased proportionally. However, intense laser pulses possess a small focal volume only ($V_f \sim 10^{-10}$\,cm$^3$, to give a typical number), while on the other side it is difficult to produce exotic atoms at very high density. An achievable number density of trapped muonic atoms is $n\sim 10^{10}$\,cm$^{-3}$ which is comparable with the densities available for other exotic species such as positronium (where $n\sim 10^{15}$\,cm$^{-3}$ \cite{Ps}) or antihydrogen (where $n\sim 10^6$\,cm$^{-3}$ \cite{antiH}). According to these numbers, only a few muonic atoms are contained in the interaction volume, which prevents a substantial yield enhancement, unfortunately.
Instead of using a fixed target of muonic atoms in a trap, it might therefore be more promising to employ a nonrelativistic beam of muonic atoms. Such beam experiments are in principle feasible and have recently led to the observation of the Ramsauer-Townsend effect in scattering of muonic hydrogen isotopes, for example \cite{Ramsauer}. Beams of 10$^5$ muons per second can be produced today \cite{muonbeam} which could be converted into the same number of muonic atoms assuming 100\% conversion efficiency. The atomic beam could be synchronized with a bunch of laser pulses: at the upcoming XFEL facilities, pulse repetition rates of 40\,kHz\,$\sim 10^{5}$\,s$^{-1}$ are envisaged \cite{XFEL}. In this setup, one muonic atom would interact with one laser pulse at a time. By assuming the highest nuclear excitation probability of about 10$^{-14}$ shown in Fig.~4, we obtain a total yield estimate of roughly one excitation event per week. This clearly indicates that an experimental observation of the NEC$\mu$M process perhaps is not completely impossible, but certainly an extremely challenging task. The signature for exciting a nucleus would be its delayed $\gamma$-emission.
Moreover, since the NEC$\mu$M probability is very low, background processes may also become relevant. In particular, it is known that the presence of charged particles in ultrastrong laser fields can give rise to nonlinear QED effects like $e^+e^-$ pair creation \cite{Burke}. They become appreciable when the laser field strength approaches the Schwinger limit $E_S=1.3\times 10^{16}$V/cm \cite{XFEL}. Since $E_{\rm OBI}\lesssim E_S$ for muonic $^{19}$F and $^{16}$N, the probability for pair creation $P_{e^+e^-}\sim \exp(-\pi E_S/E_0)$ is non-negligible at the borderline of the applicability condition $E_0\ll E_{\rm OBI}$ of our approach.

%
%
Finally, we compare the NECEM/NEC$\mu$M process with other excitation mechanisms of the nucleus via coupling with atomic states. Most closely related is nuclear excitation in laser-driven recollisions which has been considered recently \cite{recollision1,recollision2,Mocken}. Recollisions occur when the applied laser field is strong enough to tunnel-ionize the atom. The liberated electron gains energy during propagation in the field and is driven back to the nucleus when the oscillating field has reversed its direction. Upon recollision, various atomic processes can occur but also nuclear excitation. The recollision-induced excitation probability in electronic $^{239}$Pu ($\hslash\omega_N\approx 7.9$\,keV) was found to be $P\sim 10^{-16}$ at an optical laser intensity of $\sim 10^{17}$\,W/cm$^2$ \cite{recollision2}. It is larger than the NECEM probabilities found here. The reason is that the laser field can couple more effectively to a free electron, transfering large amounts of energy to it. In the above example, the electron is accelerated to weakly relativistic energies of about 10\,keV. Upon the energetic recollision, the nucleus is excited by electron scattering. As a variant of the recollision scheme, it is also conceivable to trigger nuclear transition during an OBI process where the electron or muon is violently ionized during the first laser cycle and the nucleus is excited by the resulting electromagnetic kick of the rapidly departing particle \cite{Mocken}. The circumstance that free electrons are more efficient for nuclear excitation was also observed in \cite{BeGoWe1991} in a hypothetical scenario (see Fig.\,7 therein). Similarly, the early papers on NECEM where the electrons were treated as free particles in a first approach \cite{SoBi1988}, obtained large nuclear excitation probabilities (up to $P\sim 0.1$). Such high NECEM probabilities, however, were not confirmed by experiments \cite{BoDy1992}. More efficient nuclear excitation mechanisms are resonant processes such as (field-free) NEET. They require, however, an atomic inner-shell vacancy which is usually produced by x-ray irradiation first. Contrary to that the NECEM/NEC$\mu$M process employs atoms in the ground state. In ordinary atoms the NEET probability typically amounts to $P\sim 10^{-7}$ per K-shell vacancy \cite{NEETexp,NEETexp2}. In muonic atoms the corresponding probability is largely enhanced to $P\sim 0.1$ \cite{muonicNEET2}.
It is interesting to note that coherent nuclear (or atomic) excitation in periodic fields can also occur when a fast ion beam is channeling through a crystal. For certain ion velocities a resonance behaviour arises here ('Okorokov effect') \cite{Okorokov}.

%
\section{Conclusion and Outlook}
Nuclear excitation by the nonresonant coherent motion of a strongly laser-driven bound muon has been considered. The process is interesting as it belongs to a class of excitation mechanisms which rely on the coupling of the nucleus with atomic states. By analytical means, a simple expression has been derived which provides an order-of-magnitude estimate for the total probability of electric multipole transitions in light muonic isotopes. For a given nucleus, the oscillating muon can couple much more effectively to nuclear transitions than a bound electron could, because of the smaller muonic orbital radius. In the regime of interaction considered here the absolute values of the excitation probability are still very small since the transition energies in low-$Z$ nuclei are orders of magnitude larger than the driving laser frequency. The NEC$\mu$M process therefore cannot compete with resonant excitation mechanisms like muonic NEET, for example. An experimental observation might be possible in intense x-ray laser pulses but the detection is rendered difficult by the small muonic atom densities available. Heavy isotopes with very low-lying levels in the eV range represent more promising targets for experimental studies of laser-nucleus coupling. Heavy muonic atoms are not suitable, though, since the tightly bound muon is prohibitorily difficult to influence by laser radiation. Rather in few-electron ions or bare nuclei of high $Z$, a strong, even resonant interaction with the laser field could be achieved, for example by pre-accelerating a beam of $^{235}$U ions to relativistic speed at GSI. In the case of $^{229}$Th with $\hslash\omega_N=7.6$\,eV \cite{Matinyan,Th229}, also fixed-target experiments are conceivable \cite{Habs}.

%
\section*{Acknowledgments}
A. Shahbaz acknowledges support by the Higher Education Commission (HEC), Pakistan and Deutscher Akademischer Austauschdienst (DAAD), Germany. The authors thank K. Z. Hatsagortsyan for useful discussions.

%

\end{document}